# Cosmological model in 5D: Stationarity, yes or no

W. B. Belayev[*]

*Center for Relativity and Astrophysics, box 137, 194355, Sanct-Petersburg, Russia*

The tired-light cosmology is considered in the framework of Kaluza-Klein theory in 5D. The solution of the five-dimensional semi-classical Einstein equations with nonzero five-dimensional energy-momentum tensor gives density of matter in the Universe well conformed to the observations. Variation of the light velocity and change of the rest energy and mass are interrelated. Variation of the Planck constant and electron charge is determined from formula for hydrogen spectral frequencies and observations of the fine-structure. Physical constants variation presents as explanation of anomalous change of the length of a received wave detected during radiometric analysis of Pioneer 10/11 spacecraft data. Contemporary measurements accuracy of the Microwave Cosmic Background doesn't allow to determine tendency of its parameters change permitting choice between stationary and expanding Universe's model.

PACS numbers: 98.80.-k, 04.50.+h, 04.80.Cc

## I. INTRODUCTION

The notion that the cosmological redshift is non-Doppler phenomenon arose soon after appearance Friedman-Robertson-Walker cosmology. This idea was first suggested by Zwicky [1]. Theory, explaining the redshift of the spectra of distant galaxies with variation of time scale factor was put forward by Milne [2]. He also thought the dependence of gravity constant on time $t$ to be possible [3]. Dirac [4] considered gravity constant to be inversely proportional to $t$, and time dependence of other fundamental constants. Change of mass proportional $t^2$ as a cause of decrease of the spectral frequencies of atoms is suggested in [5]. Various aspects of cosmology with non-Doppler redshift interpretation or ¡tired-light cosmology¿ are in [6]. However, with the constant length scale factor these theories can't explain the proximity of the density of matter in the Universe to critical value, which follows from the FRW model. At the same time, according to different estimates [7] the cosmological density does not exceed 1/3 of this value.

In this connection Kaluza-Klein theory is of interest. Internal space, if it exists, is considered to be forming an extra dimension in five-dimensional space-time. In [8,9] the fifth dimension induces an effective energy-momentum tensor, corresponding to the standard Einstein equations, and time scale factor is variable. The Casimir effect gives the nonzero energy-momentum tensor for the five-dimensional semi-classical Einstein equations [10]. In Sec. III we demonstrate that equations of the first type can be transformed to the equations of the second type in case of the specific metric and energy-momentum tensor, which yield cosmological model considered in Sec. VII.

Models with extra dimensions and possibilities of variation of existing constants are analyzed in [11,12]. Variation of bare "constants" of nature is assumed to be caused by variation of scale factor of internal spaces $R$. Presented theories give estimate of rate of relative changing of $R$ orders of magnitude smaller than Hubble constant $H$. However, as pointed out Barrow [12], these results base on variation of only several constant, leaving other unchanged, and if constants vary simultaneously, these limits have no real basis.

With constant fifth coordinate presented metric reduces to four-dimensional metric with cosmic time. In this frame variation of the some physical constants is obtained in Sec. IV,V. In Sec. VI we discuss use of these results for explanation of anomalous change of the length of a received wave detected during radiometric analysis of Pioneer 10/11 spacecraft and consider data of the Viking space probe radio-ranging measurements of Mars and Earth orbital radii in the same frame.

## II. EQUATIONS IN 5D

Let us show how the five-dimensional semi-classical Einstein equations with zero energy-momentum tensor are transformed to analogous equations with nonzero five-dimensional energy-momentum tensor. We consider the field

---

[*]Electronic address: vladter@ctinet.ru

equations in five dimensions [8]:

$$\hat{R}^{\alpha\beta} - \frac{1}{2}g^{\alpha\beta}\hat{R} = 0, \alpha, \beta = 0...4, \qquad (1)$$

where $\hat{R}^{\alpha\beta}$ is the Ricci tensor, $\hat{R}$ is the scalar curvature of space, $g^{\alpha\beta}$ is the metric tensor components. We look for solution in the form

$$ds^2 = N^2(t,\psi)dt^2 - \frac{1}{c_0^2}\left[dx^{1\,2} + dx^{2\,2} + dx^{3\,2} + R^2(t,\psi)d\psi^2\right], \qquad (2)$$

where $t$ is time in scale, rigid set by spectra frequencies of atoms, $c_0$ is the light velocity in a moment of time $t_0$, $x^1$, $x^2$, $x^3$ are the space coordinates of the four-dimensional space-time, $\psi$ is the coordinate of extra dimension, created by compactified internal space, $N(t,\psi)$, $R(t,\psi)$ are the arbitrary functions $t$ and $\psi$. Let's denote overdot the derivative with respect to $t$, derivative with respect to $\psi$ by $(*)$ and $x^0 \equiv t$, $x^4 \equiv \psi$. Then the nonzero components of a tensor $\hat{R}^{\alpha\beta}$ are

$$\hat{R}^{00} = -\frac{\ddot{R}_0}{R_0 N_0^4} + c_0^2 \frac{N_0^{**}}{R_0^2 N_0^3} + \frac{\dot{N}_0 \dot{R}_0}{R_0 N_0^5} - c_0^2 \frac{N_0^* R_0^*}{R_0^3 N_0^3}, \qquad (3)$$

$$\hat{R}^{44} = -c_0^4 \frac{N_0^{**}}{N_0 R_0^4} + c_0^2 \frac{\ddot{R}_0}{N_0^2 R_0^3} + c_0^4 \frac{R_0^* N_0^*}{N_0 R_0^5} - c_0^2 \frac{\dot{R}_0 \dot{N}_0}{N_0^3 R_0^3}, \qquad (4)$$

where $N_0$, $R_0$ are coefficients of metric (2) being solution of Eq. (1). Value of the scalar curvature of the space turns out to be

$$\hat{R} = g_{00}\hat{R}^{00} + g_{44}\hat{R}^{44} = -2\frac{\ddot{R}_0}{R_0 N_0^2} + 2c_0^2 \frac{N_0^{**}}{N_0 R_0^2}. \qquad (5)$$

The nonvanishing equations (1) are

$$\frac{\dot{N}_0 \dot{R}_0}{R_0 N_0^5} - c_0^2 \frac{N_0^* R_0^*}{R_0^3 N_0^3} = 0$$

$$c_0^2 \frac{\ddot{R}_0}{R_0 N_0^2} - c_0^4 \frac{N_0^{**}}{N_0 R_0^2} = 0 \qquad (6)$$

$$c_0^4 \frac{R_0^* N_0^*}{N_0 R_0^5} - c_0^2 \frac{\dot{R}_0 \dot{N}_0}{N_0^3 R_0^3} = 0.$$

We rewrite Eq. (1) as

$$\hat{R}^{\alpha\beta} - \frac{1}{2}g^{\alpha\beta}\hat{R} + 8\pi G T^{\alpha\beta} = 8\pi G T^{\alpha\beta}, \alpha, \beta = 0...4, \qquad (7)$$

where $T^{\alpha\beta}$ is the five-dimensional energy-momentum tensor, $G$ is gravitational constant slowly varying with $t$.

We consider a five-dimensional energy-momentum tensor with nonzero diagonal matrix elements look like

$$T_i^j = diag\left(\rho, \frac{p_{ext}}{c^2}, \frac{p_{ext}}{c^2}, \frac{p_{ext}}{c^2}, \frac{p_{int}}{c^2}\right), \qquad (8)$$

where $c$ is light velocity varying with $t$, $\rho$ is the density of matter in the Universe, $p_{ext}$ is the isotropic pressure in external three-dimensional space, $p_{int}$ is a parameter with dimension, which equals to the pressure. We write the energy-momentum tensor as

$$T^{ij} = \left(\rho + \frac{p^{ij}}{c^2}\right)u^i u^j - g^{ij}\frac{p^{ij}}{c^2}, \qquad (9)$$

where $p^{ij} = 0$, $i \neq j$; $p^{ii} = p_{ext}$, $i = 0...3$; $p^{44} = p_{int}$, and $u^i = dx^i/ds$ are the components of velocity vector of the fluid elements. Assuming $u^i = 0$, $i = 1...4$, from metric (2) we obtain

$$u^0 = N^{-1}. \qquad (10)$$

The nonzero components of the energy-momentum tensor is presented as

$$T^{00} = \rho N^{-2}$$

$$T^{ii} = c_0^2 c^{-2} p_{ext}, \ i = 1, 2, 3 \tag{11}$$

$$T^{44} = c_0^2 c^{-2} p_{int} R^{-2}.$$

If we assume $N_0$ and $R_0$ independent of $\psi$, then taking into account values of elements of tensor $T^{\alpha\beta}$ (11) equations (6) in form (7) give

$$\frac{\dot{N}_0 \dot{R}_0}{R_0 N_0^3} + 8\pi G \rho = 8\pi G \rho$$

$$\frac{\ddot{R}_0}{R_0 N_0^2} + 8\pi G c^{-2} p_{ext} = 8\pi G c^{-2} p_{ext} \tag{12}$$

$$-\frac{\dot{R}_0 \dot{N}_0}{N_0^3 R_0} + 8\pi G c^{-2} p_{int} = 8\pi G c^{-2} p_{int}.$$

Let's set functions $N(t)$, $R(t)$ by equalities

$$\frac{\dot{N}\dot{R}}{RN^3} = \frac{\dot{N}_0 \dot{R}_0}{R_0 N_0^3} + 8\pi G \rho$$

$$\frac{\ddot{R}}{RN^2} = \frac{\ddot{R}_0}{R_0 N_0^2} + 8\pi G c^{-2} p_{ext} \tag{13}$$

$$-\frac{\dot{R}\dot{N}}{N^3 R} = -\frac{\dot{R}_0 \dot{N}_0}{N_0^3 R_0} + 8\pi G c^{-2} p_{int}$$

which constrain other parameters in these equations. Substituting values of the right parts of equalities (13) in Eq. (12) we obtain

$$\frac{\dot{N}\dot{R}}{RN^3} = 8\pi G \rho \tag{14}$$

$$\frac{\ddot{R}}{RN^2} = 8\pi G c^{-2} p_{ext} \tag{15}$$

$$\frac{\dot{R}\dot{N}}{N^3 R} = -8\pi G c^{-2} p_{int} \tag{16}$$

which are reduced from equations

$$\hat{R}^{\alpha\beta} - \frac{1}{2} g^{\alpha\beta} \hat{R} = 8\pi G T^{\alpha\beta}, \alpha, \beta = 0...4, \tag{17}$$

with energy-momentum tensor (9) and for metric (2). Thus, the five-dimensional semi-classical Einstein equations with zero energy-momentum tensor (1) is allowed to transform for this metric to Eq. (17) with nonzero energy-momentum tensor (9).

### III. METRIC FOR TIRED-LIGHT THEORY IN 4D

The metric (2) with $\psi = const$ reduces to four-dimensional metric

$$ds^2 = c_0^2 N^2(t) dt^2 - dr^2 + r^2 d\theta^2 + r^2 \sin^2\theta d\phi^2, \tag{18}$$

where $r$, $\theta$, $\phi$ are the spherical coordinates, $N$ is the coefficient dependent on $t$. This metric with constant $N(t)$ turns to Minkowsky space-time. Assuming $N(t)$ to be slow changing and substituting

$$c = c_0 N(t) \tag{19}$$

we get the standard Minkowsky metric. Therefore, light velocity in four-dimensional coordinate system with its center in $(t, O)$, where $O$ is an arbitrary point moving without action of any forces in three-dimensional space is represented by (19). Let us denote

$$\tau = \int_0^t N(t)dt \tag{20}$$

as a time in $(t_0, O)$ coordinate system. The expression connected a distance intervals $dr_0$ in $(t_0, O)$ system and $dr$ in $(t, O)$ system follows from metric (18):

$$dr_0 = dr. \tag{21}$$

Then, a velocity $v_0$ in $(t_0, O)$ coordinate system could be represented as

$$v_0 = \frac{dr_0}{d\tau}. \tag{22}$$

The energy of a particle for the metric (18) is [13]:

$$E = \frac{m_0 c_0^2 N(t)}{(1 - v_0^2/c_0^2)^{\frac{1}{2}}}, \tag{23}$$

where $m_0$ is the rest mass of the particle in $(t_0, O)$ coordinate system. Assuming $v_0 = 0$, we get expression for rest energy of the particle

$$E = m_0 c_0^2 N(t). \tag{24}$$

Hence

$$E = E_0 N(t), \tag{25}$$

where $E_0 = m_0 c_0^2$ is the rest energy of the particle in coordinate system with its center in $(t_0, O)$. In view of (19) from (24) we obtain $E = m_0 c^2 / N(t)$. From this it follows that the rest mass is changing with time as

$$m(t) = \frac{m_0}{N(t)}. \tag{26}$$

This variation of mass is relative, i.e, does not occur as variation of mass as number of nucleons.

Let us determine the magnitude of force acting upon the particle. Expression for the vector of the force at small velocities in relation to coordinate system $(t, O)$ is represented by $F = mdV/dt$, where $V$ is the velocity vector. Since (22) the velocity vector in coordinate system $(t_0, O)$ is $V_0 = V/N(t)$, because of (26), (20) the vector of the force is written as

$$F = \frac{m_0}{N(t)} \frac{d(V_0 N(t))}{dt} = \frac{m_0}{N(t)} \left[ \dot{N}(t) V_0 + \frac{dV_0}{d\tau} N^2(t) \right]. \tag{27}$$

Regarding $\dot{N}$ to be small, we get the magnitude of force, acting upon the particle in system $(t, O)$:

$$f = m_0 \frac{dv_0}{d\tau} N(t) = f_0 N(t), \tag{28}$$

where $f_0$ is its value in system $(t_0, O)$.

## IV. REDSHIFT AND PHYSICAL CONSTANTS

We determine the redshift magnitude considered tired-light model as

$$z = \frac{(\lambda_0(\tau) - \lambda_0(\tau_0))}{\lambda_0(\tau_0)}, \tag{29}$$

where $\lambda_0(\tau)$ is the wavelength of emission, radiated in moment $\tau$ in system $(t_0, O)$, $\tau_0 = \tau(t_0)$, where $t_0$ is the present time moment. Because light velocity $c_0$ in this coordinate system is constant in any point of the space-time, this formula gives

$$z = \frac{(P_0(\tau) - P_0(\tau_0))}{P_0(\tau_0)}, \tag{30}$$

where $P_0(\tau)$ is period of emission in system $(t_0, O)$. Since $P_0(\tau) = PN(t)$, where $P$ is period of emission in system $(t, O)$, which is constant at any system, then in view of $N(t_0) = 1$ we have

$$z = N(t) - 1. \tag{31}$$

Assuming the rate of change $N(t)$ to be constant in time interval $\Delta t = t_0 - t$, we write $z = -\dot{N}(t_0)\Delta t$. Denoting

$$H = -\frac{\dot{N}(t_0)}{N(t_0)}, \tag{32}$$

we obtain Hubble Law $zc_0 = H\Delta r_0$, where $\Delta r_0 = \Delta t_0 c_0$.

The variation of light velocity per time unit is determined from (19):

$$\dot{c}(t_0) = c_0 \dot{N}(t_0) = -c_0 H. \tag{33}$$

Assuming the Hubble constant $H = 65$ km s$^{-1}$ Mpc$^{-1}$ [14], we will get $\dot{c}(t_0) = 1.95$ cm s$^{-1}$ yr$^{-1}$. The light velocity is changed in time scale set by atoms frequencies, which correspond coordinate system $(t, O)$.

Relation between the gravity constant $G_0$ in system $(t_0, O)$ and its value $G$ in system $(t, O)$ determines from Newton Law $d^2 r/dt^2 = -GM/r^2$, where $M$ is spherical mass in system $(t, O)$. Because of mass change (26), invariable distance interval (21) and relation $d\tau = N(t)dt$ we obtain

$$G = G_0 N^3. \tag{34}$$

Disregarding possible variation $G_0(\tau)$ with time, orbit of the point, moving in gravity field, doesn't change within the $(t_0, O)$ coordinate system.

Let us consider variation of Planck constant $h$ with time. The quantum energy in system $(t, O)$ is $E_q = h/P$, where $P$ is oscillation period of the light wave. In accordance with (25) in coordinate system $(t_0, O)$ in time moment $\tau(t)$ this energy is

$$E_{q0}(\tau) = E_q N^{-1}(t). \tag{35}$$

Question, is the absorption energy of free quantum changing the same as rest energy of the particle with mass is not evident. We write

$$E_q = E_{q0}(\tau_0)\Phi(t)N(t), \tag{36}$$

where $\Phi(t)$ is function $t$. Because the oscillation period of free quantum in the moment $t_0$ within coordinate system $(t_0, O)$ in accordance with (20) is

$$P_0 = PN(t), \tag{37}$$

Eq. (36) yields

$$h(t) = h_0 \Phi(t), \tag{38}$$

where is $h_0$ is the Planck constant in time moment $t_0$ in $(t_0, O)$ coordinate system. Since in this coordinate system in time moment $\tau$ period is $P_0(\tau) = P_0$, in accordance with (35) and (37) we have

$$h_0(\tau(t)) = h(t). \tag{39}$$

Variation of the charge of the electron $e$ is obtained from formula of hydrogen spectral frequencies $\nu_{kn}$ for system $(t, O)$:

$$\nu_{kn} = \frac{2\pi^2 m_e e^4}{h^3}\left(\frac{1}{k^2} - \frac{1}{n^2}\right), \tag{40}$$

where $m_e$ is electron mass, $k, n$, $n > k$ are natural numbers. Since spectral frequencies set time scale in system $(t, O)$, $\nu_{kn}$ is constant. Taking into account change of the electron mass (26) and Planck constant (39) Eq. (40) gives

$$e = e_0 \Phi^{\frac{3}{4}}(t) N^{\frac{1}{4}}(t), \tag{41}$$

where $e_0$ is charge of electron in system $(t_0, O)$ in time moment $t_0$. On the other hand, in system $(t_0, O)$ spectral frequencies change as $\nu_{kn}(\tau) = \nu_{kn}/N(t)$, and electron mass doesn't vary. Then Planck constant variation (39) and formula (40) for electron charge in time moment $\tau$ in this coordinate system yield

$$e_0(\tau) = e_0 \Phi^{\frac{3}{4}}(t) N^{-\frac{1}{4}}(t). \tag{42}$$

Function $\Phi(t)$ is determined by means of fine-structure $\alpha$. Its observed values $\alpha_0(\tau) = e_0(\tau)^2/(h_0(\tau)c_0)$ slow change with $z$ [15]. Hence, in the first approximation from Eqs. (39), (38) and (42) we obtain

$$\Phi = N. \tag{43}$$

Therefore, the absorption energy of free quantum $E_q$ (36) changes with time as

$$E_q = E_{q0} N^2(t), \tag{44}$$

Planck constant (38) varies as

$$h = h_0 N(t), \tag{45}$$

and electron charge (41), (42) varies as

$$e = e_0 N(t) = e_0(\tau) N^{\frac{1}{2}}(t). \tag{46}$$

## V. ENERGY PROCESSES AND GENERATION OF MATTER

We define the energy change (24) with time

$$W = \frac{dE}{dt} = m_0 c_0^2 \dot{N}(t). \tag{47}$$

In view of (19) and (26) we can write $W = \frac{\dot{N}}{N} mc^2$, and, as follows from (32), at present time

$$W = -H m_0 c_0^2. \tag{48}$$

Thus, energy of a body with mass 1 kg decreases by 0.186 J every second. Since, as follows from energy conservation law, it doesn't disappear, it means that the energy is liberated in some way. However, according to (48) the Sun should liberate $3.7 \times 10^{29}$ J/sec, while it actually does $3.8 \times 10^{26}$ J/sec. Therefore, it is allowed to suppose that this energy is consumed for matter creation. Bondi and Gold [16], Hoyle [17], Jordan [18] put forward hypothesis of the matter creation in the framework of expanding Universe model. Analysis of currents of helium, coming out of depths of the Earth, testifies about its radiogenic origin [19]. On the basis of these data, the hypothesis of generation of matter within the Earth gains further development [20]. However, to prove this notion, additional arguments are necessary.

Let's determine, what quantity of matter could emit in the form of additional nucleons in a unit of time, if all energy (47) is used for its generation. We consider the change of mass $m_0(t)$ resulting from generation of matter in moment of time $\tau(t)$ in coordinate system $(t_0, O)$. This mass emits per one time unit (47):

$$U = -W = -\dot{N}(t) m_0(t) c_0^2. \tag{49}$$

On the other hand, formula (24) yields

$$U = \dot{m}_0(t) N(t) c_0^2. \tag{50}$$

where $\dot{m}_0$ is the mass, generated per time unit in system $(t_0, O)$. Solving equation $\dot{m}_0(t)/m_0(t) = -\dot{N}(t)/N(t)$ with initial conditions $m_0(t_0) = 1$, $N(t_0) = 1$, we obtain

TABLE I. Values of Hubble constant, coefficients $A$, following from Pioneer 10/11 data, $B_E$ for Earth, $B_M$ for Mars, following from Viking data.

| $A$ | $H$ | $B_E$ | $B_M$ |
|---|---|---|---|
| 1 | 43.7 | 1.8±0.9 | 5.7±0.6 |
| 0.5 | 65.6 | 1.2±0.6 | 3.9±0.4 |
| 0 | 87.4 | 0.9±0.45 | 2.9±0.3 |

$$m_0(t) = \frac{m_0}{N(t)} \tag{51}$$

and, at present time,

$$\dot{m}_0(t_0) = H m_0. \tag{52}$$

It should be pointed out, however, it is not essential, that all emitted energy is used for generation of matter, but in several cases, probably, is released in some other forms, for instance, as heat.

## VI. ANALYSIS OF THE TRAJECTORIES IN SOLAR SYSTEM

Radiometric analysis of Pioneer 10/11 spacecraft data [21] indicate additional acceleration $a_r \approx 8.5 \times 10^{-8}$ cm/s$^2$, directed towards the Sun. Detected change of the length of a received wave $\lambda$ can be a result of two factors effect: decrease of the light velocity and variation of the cycle of radiation of the spacecraft generator $P_g$ in the scale of receiving equipment

$$\dot{\lambda} = \dot{c} P_g + c \dot{P}_g . \tag{53}$$

Variation of the cycle of radiation might be caused by dependence of processes, which determine work of generator and receiving equipment, on time scale factor because of physical constants variation. Considering $P_g(t) = \hat{P}(N(t))$ we obtain

$$\dot{P}_g(t_0) = -H \frac{d\hat{P}(N(t_0))}{dN}. \tag{54}$$

Denoting $A = \frac{d\hat{P}(N(t_0))}{dN}$ and taking into account change of the light velocity (33) we can write Eq. (53) as

$$\dot{\lambda}(t_0) = -(1+A) H c_0 P_g(t_0). \tag{55}$$

The rate of the relative change of wavelength is

$$\frac{\dot{\lambda}(t_0)}{\lambda} = -(1+A) H. \tag{56}$$

At the same time it is presented as $\dot{\lambda}/\lambda = a_r/c$, in assumption cycle of radiation and light velocity are constant, and variation of wavelength is caused by additional acceleration. Some values of $A$ and $H$ is in Table I.

Viking space probe to Mars provided radio-ranging measurements to an accuracy of 12 m [21]. Their data determines the change difference between the Mars and Earth orbital radii about 100 m in synodic period, and the change their sum about 150 m in synodic period. Let us determine possible relation between change orbital radii and $H$ from equation of the point motion in central gravity field in 3D, considering force to be depending on $N(t)$ :

$$\ddot{r} - r\dot{\gamma}^2 = -\frac{GM}{r^2} Y(N(t)), \tag{57}$$

where $\gamma$ is angular coordinate, $M$ is mass of the Sun in coordinate system $(t, O)$ disregarding possible mass generation, $Y$ is function of $N(t)$. For small interval $\Delta t = t - t_0$ we have

$$Y(N(t)) = 1 + \dot{N}(t_0) \frac{dY(N)}{dN}(t_0) \Delta t = 1 + H B \Delta t, \tag{58}$$

where $B = -\frac{dY(N)}{dN}(t_0)$. Eq. (57) with the areas integral $r^2\dot{\gamma}/2 = C$, where $C$ is constant, gives

$$\ddot{r} = \frac{4C^2}{r^3} - \frac{GM}{r^2}(1 + BH\Delta t).  \qquad (59)$$

Since $d\tau = N(t)dt$, with use (26), (34) and without small quantity of higher order, in coordinate system $(t_0, O)$ this equation is rewritten as

$$\frac{d^2 r}{d\tau^2} = \frac{4C_0^2}{r^3} - \frac{G_0 M_0}{r^2}(1 + BH\Delta\tau), \qquad (60)$$

where $C_0 = C/N(t)$. In linear form equation of motion yields

$$\frac{d^2 r}{d\tau^2} = \frac{4C_0^2}{r_b^3} - \frac{G_0 M_0}{r_b^2} - \left[\frac{12 C_0^2}{r_b^4} + \frac{2 G_0 M_0}{r_b^3}\right]\Delta r \\ - \frac{G_0 M_0}{r_b^2} BH\Delta\tau + O((\Delta r)^2) + O(\Delta\tau\Delta r), \qquad (61)$$

where $\tau_b$ is initial time, $r_b$ is initial distance between mass point and center of gravity, $\Delta r$ is its change. Considering the case of a circular orbit $2C_0 = \sqrt{G_0 M_0 r_b}$, we make assumption $d^2 r/d\tau^2 = 0$, whose rightness is confirmed by numerical solution of Eq. (60) with use Runge-Kutta method. Value of the velocity asymptotically approaches to solution of Eq. (61):

$$\frac{dr}{d\tau} = -HBr_b. \qquad (62)$$

Values of coefficient $B$ for Earth and Mars, following from Viking data, are in Table I. Possible increment mass of the Sun caused of its emission as a result of decrease the rest energy of the particles through change time scale factor (52) gives $B = 1$. In the Kaluza-Klein and superstring theories it is predicted, that gravitational constant inversely proportional to the volume of internal space [12,11]. In [22,23] change of gravity constant is considered as a cause of the Earth and Mars orbit radii decrease. Assumption of the relative change of the mass with time, but without particles generation, in frame of 4+1 Kaluza-Klein theory for Big Bang model with changing time scale factor [9,8] is the other approach to this matter. In this connection, it should note, however, striking proximity values $BH$ to relative rate of a deceleration of the Earth rotation $\approx 2.37 \times 10^{-10}$ year$^{-1}$ [24].

## VII. COSMOLOGICAL PARAMETERS

The mean value of density of matter in the Universe, obtained by different techniques [7] with error $50-75\%$, comprises about $1/5$ of critical density, defined with help of the Fridman-Robertson-Walker model, $\rho_{crit} = 3H^2/(8\pi G_0)$. Assuming that

$$\dot{N}(t_0) = \dot{R}(t_0) \qquad (63)$$

we obtain from Eq. (14), using (32), at $t = t_0$ the density value

$$\rho_0 = \frac{1}{8}\frac{H^2}{\pi G_0}. \qquad (64)$$

which is within the limits of the density of matter in the Universe, derived from measurements. It is natural to assume, that if equality of rate of change of the time scale factor and length of internal space (63) takes place at present, it is fulfilled permanently.

Function $N(t)$, determining redshift, is found from equation obtained by substitution following from (63) relation $R = N$ in Eq. (14):

$$\frac{\dot{N}^2}{N^4} = 8\pi G\rho, \qquad (65)$$

taking into account possible variation of $G_0(\tau)$ with time and generation of matter considered in Sec. V.

Eqs. (14)-(15 give equations of state

$$p_{int} = -c^2\rho \tag{66}$$

$$\ddot{\Pi}_{int}\Pi_{int}(p_{ext} + 2p_{int}) = \left(\dot{\Pi}_{int}\right)^2(p_{ext} + 5p_{int}) + \dot{\Pi}_{int}\dot{\Pi}_{ext}p_{int},$$

where $\Pi_{int} = Gc^{-2}p_{int}$, $\Pi_{ext} = Gc^{-2}p_{ext}$.

Because of variation Planck constant (45) photons absorption energy of the Microwave Cosmic Background $E_q$ change gives by (44). Figure 1. shows variation of blackbody spectrum for the MCB in case of tired light model and FRW cosmology. At present time the highest relative accuracy of measurement of the MCB intensity is less than $10^{-6}$ [25]. For determination of nature of the intensity change during 10 years the relative accuracy above $10^{-9}$ is necessary. In considered variant of the tired-light theory, just as in the Big-Bang model, MCB is residual after recombination. Two versions must be analysed:

1). The Universe was arisen by means of the explosion and then expansion finished.

2). The Universe was stationary during whole time. Let's consider in this connections Eq. (65). If we denote $X(t) = \rho_0(\tau(t))G_0(\tau(t))/(\rho_0 G_0)$, where $\rho_0(\tau)$ is variable density in coordinate system $(t_0, O)$, then using (26), (34) and taking into account value of H (64) we obtain

$$\frac{\dot{N}^2}{N^6} = H^2 X(t). \tag{67}$$

With variable $\tau$ Eq. (67) is rewritten as

$$\frac{1}{N^4}\left(\frac{dN}{d\tau}\right)^2 = H^2 X(t(\tau)). \tag{68}$$

This equation gives

$$\frac{1}{N^2 X^{\frac{1}{2}}}\frac{dN}{d\tau} = H. \tag{69}$$

The solution of Eq. (69) is

$$I(N) \equiv \int_{N(t(\tau_0))}^{N(t(\tau))} \frac{1}{N^2 X^{\frac{1}{2}}} dN = H(\tau - \tau_0). \tag{70}$$

If $I(N) \to \infty$ with $N \to \infty$, then this equation doesn't give limitation for $\tau_0$. This condition is fulfilled, in particular, when

$$\frac{1}{X} = O(N^2), \tag{71}$$

or $\frac{1}{X}$ is variable quantity higher order. In this case certain processes of nuclear synthesis might run on in rare plasma with gradual decrease of the photon energy in the course of estimably more protractedly in coordinate system $(t_0, O)$ than if Big Bang took place. This model admits, that the time of Universe's existence until recombination was being more than after its. Therefore, photons in visible field had interaction until recombination, that gives homogeneity of the MCB angular power spectrum. Still, theory of depending on the time the elementary particles properties is necessary for basing or refutation of this notion, and this one isn't only problem, which must be solved to give reasons for such model.

The FRW model explains observed time dilation of high redshift supernova [26] by Doppler effect. Consideration of dependence of the supernova processes upon variation physical constant is a possible way for explanation of this phenomenon in the framework of the tired-light theory.

## VIII. CONCLUSION

The tired-light theory didn't gain such development as the FRW cosmology. However, we judge of the Universe's expansion by indirect data. Some of them, such as ratio of hydrogen and helium, are well explained by the FRW model. For explanation other of them, including anomalous change of the length of a received radio-wave from Pioneer 10/11, the tired-light theory has more means. Therefore, this model merit the same close attention as FRW cosmology.

## ACKNOWLEDGMENTS

The author would like to thank A.I. Cigan of the Ioffe Physical-Technical Institute for valuable discussion. I thank Slava G. Turyshev for information about Viking mission.

[1] F. Zwicky, Proc. Nat. Acad. Sci. USA **15**, 773 (1929).
[2] E.A. Milne, *Relativity, Gravitation and World-Structure* (Oxford University Press, Oxford, 1935).
[3] E.A. Milne, Proc. R. Soc. London Ser. A **158**, 324 (1937).
[4] P.A.M. Dirac, Nature **139**, 323 (1937).
[5] J. Narlikar, H. Arp, Astrophus. J. **405**, 51 (1993)
[6] P.A. LaViolette, Astrophus. J. **301**, 544 (1986)
[7] N.A. Bahcall and X. Fan, Proc. Nat. Acad. Ski. USA **95**, 5956 (1998) (also astro-ph/9804082).
[8] J.M. Overduin and P.S. Wesson, Phus. Rept. **283**, 303 (1997) (also gr-qc/9805018).
[9] P.S. Wesson and H. Liu, Astrophus. J. **440**, 1 (1995).
[10] L.-X. Li, J.R. III Gott, Rhys. Rev. D**58**, 103513 (1998) (also astro-ph/9804311).
[11] E.W. Kolb, M.J. Perry and T.P. Walker, Phys. Rev. D**33**, 869 (1986).
[12] J.D. Barrow, Phys. Rev. D**35**, 1805 (1987).
[13] L.D. Landau, E.M. Lifshitz, *Teoriya Polja* (Nouka, Moscow, 1973).
[14] W.L. Freedman et al., Nature **371**, 752 (1994).
[15] D.A. Varshalovich and A.Y. Potekhin, Space Sci. Rev. **74**, 259 (1995).
[16] H. Bondi, *Cosmology* (Cambridge University Press, Cambridge, 1961).
[17] F. Hoyle, Nature **163**, 4136 (1949).
[18] P. Jordan, *Schwerkraft und Weltall* (Fr. Vieweg und Sohn, Braunschweig, 1952).
[19] I.N. Yanitscy, in: *Geophysica i Sovremenny Mir, Megdunarodnaya Nouchnaya Conferenciya, Doklady i Referaty* (VINITI Publ., Moscow, 1993).
[20] K.I. Sokolovsky, K.E. Yesipchuk, in: *Geophysica i Sovremenny Mir, Megdunarodnaya Nouchnaya Conferenciya, Doklady i Referaty* (VINITI Publ., Moscow, 1993).
[21] J.D. Anderson et al., Phys. Rev. Lett. **81**, 2858 (1998) (also gr-qc/9808081).
[22] C.Wetterich, Nucl. Phys. B**302**, 643 (1988).
[23] R.W.Hellings et al., Phys. Rev. Lett. **51**, 1909 (1983).
[24] A.M. Mikisha, *Cosmicheskiye metodi v geodeziyi* (Znaniye, Moscow, 1983).
[25] M. Tegmark Astrophus. J. Lett. **69**, 514 (1999) (also astro-ph/9809201).
[26] B. Leibungut et al., Astrophus. J. **L21-L24**, 466 (1996) (also astro-ph/9605134).

FIG. 1. Evolution of the MCB blackbody spectrum. Line 1 is FIRAS spectrum of MCB. The FRW cosmology (line 2) and tired light theory (line 3) predicted spectra correspond $z = 0.1$. Curves 1, 2 and data for curve 3 are from E. L. Wright, not published.